\documentclass[a4paper]{jpconf}

\usepackage[english]{babel} 

\usepackage[%
	final,%
]{graphicx}

\usepackage[
   centertags, 
   sumlimits,  
   intlimits,  
   namelimits, 
]{amsmath} %
\usepackage{amssymb}

\usepackage[%
	pdftitle={Relativistic Shock Waves and Mach Cones in Viscous Gluon Matter},%
	pdfauthor={Ioannis Bouras, Etele Moln\'ar, Harri Niemi, Zhe Xu, Andrej El, Oliver Fochler, Francesco Lauciello, Carsten Greiner and Dirk H.\ Rischke},%
	pdfsubject={Investigation of shocks in viscous matter},
	pdfstartview=FitH,
	pdfpagemode=UseNone,
	bookmarksopen=true
	]{hyperref}

\usepackage[all]{hypcap} 

\usepackage{dcolumn} 

\usepackage{multirow} 

\usepackage{microtype} 

\makeatletter
\def\tagform@#1{\maketag@@@{\ignorespaces#1\unskip\@@italiccorr}}
\let\orgtheequation\theequation
\def\theequation{(\orgtheequation)}
\makeatother

\newcommand{\beq}{\begin{equation}}
\newcommand{\eeq}{\end{equation}}

\begin{document}
\title{Relativistic Shock Waves and Mach Cones in Viscous Gluon Matter}

\author{Ioannis Bouras$^1$, Etele Moln\'ar$^{2,3}$, Harri Niemi$^2$, Zhe Xu$^{1,2}$, Andrej El$^1$, Oliver Fochler$^1$, Francesco Lauciello$^1$, Carsten Greiner$^1$ and Dirk H.\ Rischke $^{1,2}$}

\address{$^1$ Institut f\"ur Theoretische Physik, Johann Wolfgang Goethe-Universit\"at,
Max-von-Laue-Str.\ 1, D-60438 Frankfurt am Main, Germany}

\address{$^2$ Frankfurt Institute for Advanced Studies, Ruth-Moufang-Str. 1, D-60438 Frankfurt am Main, Germany}

\address{$^3$ KFKI, Research Institute of Particle and Nuclear Physics, H-1525 Budapest, P.O.Box 49, Hungary}

\ead{bouras@th.physik.uni-frankfurt.de}

\begin{abstract}

To investigate the formation and the propagation of relativistic
shock waves in viscous gluon matter we solve the relativistic Riemann
problem using a microscopic parton cascade. We demonstrate
the transition from ideal to viscous shock waves by varying
the shear viscosity to entropy density ratio $\eta/s$. Furthermore
we compare our results with those obtained by solving the relativistic causal
dissipative fluid equations of Israel and Stewart (IS), in order to show 
the validity of the IS hydrodynamics. 
Employing the parton cascade we also investigate the formation of Mach
shocks induced by a high-energy gluon traversing viscous gluon matter.
For $\eta/s = 0.08$ a Mach cone structure is observed, whereas the signal
smears out for $\eta/s \geq 0.32$.

\end{abstract}

\section{Introduction}
Jet quenching has been discovered in 
heavy-ion collisions at BNL's Relativistic Heavy Ion Collider 
(RHIC). In this context, very exciting 
jet-associated particle correlations \cite{Wang:2004kfa} 
have been observed, which indicates the formation of
shock waves in the form of Mach cones \cite{Stoecker:2004qu},
induced by supersonic partons moving through the 
quark-gluon plasma (QGP). Measuring the Mach cone angle
could give us the possibility
to extract the equation of state of the QGP.

Shock waves can only develop in a medium which behaves like a fluid.
The large elliptic flow coefficient $v_2$ measured at RHIC \cite{Adler:2003kt}
implies that the QGP created could be a nearly perfect fluid
with a small viscosity. Calculations of viscous hydrodynamics
\cite{Luzum:2008cw} and microscopic transport theory \cite{Xu:2007jv,Xu:2008av} 
have estimated the shear viscosity to the entropy density ratio
$\eta/s$ to be less than 0.4. There is still an open question if this
upper limit of the $\eta/s$ ratio is small enough to allow the formation
of Mach shocks.

In this work we address the question, whether and when relativistic
shock waves and Mach cones can develop in viscous gluon matter for given 
$\eta/s$ values. For this purpose we consider first the relativistic Riemann 
problem \cite{Schneider:1993gd}, which we solve within the kinetic theory 
and the Israel-Stewart (IS) theory of viscous hydrodynamics for comparisons.
Here we employ the microscopic parton cascade BAMPS
(Boltzmann Approach of MultiParton Scatterings) \cite{Xu:2004mz}
and a solver of the IS equations, vSHASTA 
(viscous SHArp and Smooth Transport Algorithm) \cite{Molnar:2009tx}.
Particularly, we demonstrate agreements between the two approaches for matter
with (extreme) small $\eta/s$ values and also show deviations when the
$\eta/s$ ratio becomes large, which implies the invalidity of the IS theory.
Second, we consider a traverse of a high-energy gluon through gluon matter
and investigate the formation of shocks in form of Mach cones. Preliminary
results are obtained by using BAMPS.

\section{BAMPS and vSHASTA}
\label{model}
BAMPS is a microscopic transport model solving the Boltzmann equation
\begin{equation}
p^{\mu} \partial_{\mu} f(x,p) = C(x,p)
\end{equation}
for on-shell particles with the collision integral $C(x,p)$.
The algorithm for collisions is based on the stochastic interpretation 
of the transition rate \cite{Xu:2004mz}. In this study, we consider
only binary gluon scattering processes with an isotropic
cross section, which is adjusted locally at each time
step to keep a constant $\eta/s$ 
value \cite{Bouras:2009nn,Xu:2007ns,El:2008yy}.

Simulations of particle evolution in space and time are performed
in a static box, where the whole box is divided into spatial cells with a volume
$V_{\rm cell} = \Delta x \, \Delta y \, \Delta z$. Collisions of
particles in same cells are simulated by Monte Carlo technique according
to the individual collision probability within a time step of $\Delta t$,
\begin{equation}\label{eq_BAMPS_collProb}
P_{ \rm 22} = v_{ \rm rel} \frac{ \sigma }
{ N_{\rm test} } \frac{\Delta t}{ V_{\rm cell} } \, ,
\end{equation}
where $\sigma$ is the total cross section,
$v_{ \rm rel} = (p_1 + p_2)^2/(2 E_1 E_2)$ denotes the relative
velocity of the two incoming particles with four momenta $p_1, p_2$
and $N_{\rm test}$ is the testparticle number.
The testparticle method is introduced to reduce statistical fluctuations
and is implemented such that the mean free path is left invariant.

The relationship between the shear viscosity $\eta$ and the total cross
section $\sigma$ is given by $\eta = 4 e / (15 R^{tr})$ \cite{Xu:2007ns},
where $R^{tr} = n \langle v_{\rm rel} \sigma^{tr} \rangle= 2 n \langle v_{\rm rel} \sigma \rangle /3$
is the transport collision rate in case of isotropic scattering processes.
$\langle \rangle$ stands for ensemble average in local rest frame.
We obtain
\begin{equation}\label{eq_BAMPS_shearViscosity}
\eta = \frac{2}{3} e \lambda_{\rm mfp}
\end{equation}
for an ultrarelativistic massless gas. Here the LRF energy density
is $e=3nT$, $n$ is the LRF particle density, $T$ the temperature
and $\lambda_{\rm mfp} = 1/(n \langle v_{\rm rel} \sigma \rangle)$
is the particle mean free path.
In kinetic equilibrium the LRF entropy density
is given by $s = 4 n - n \ln \lambda$, where
$\lambda = n / n_0 $ is the fugacity, the ratio of the LRF number density
to the one in thermal equilibrium $n_0 = 16 T^3 / \pi^2$.

vSHASTA solves the IS equations of dissipative
hydrodynamics for shear pressure and heat flow. In 1+1 dimensions
the relaxation equations for heat conductivity and shear stress are
\begin{eqnarray}
D q^z & = & \frac{1}{\tau_q}\left(q^z_{NS} - q^z\right)
- I^z_{q1} - I^z_{q2} - I^z_{q3}\, , \\
D\pi &=& \frac{1}{\tau_\pi}\left(\pi_{NS} - \pi \right)
- I_{\pi1} - I_{\pi2} - I_{\pi 3}\, ,
\end{eqnarray}
with
\begin{eqnarray}
I^z_{q1} &=& \frac{1}{2}q^z
\left(\theta_z + D \ln \frac{\beta_1}{T} \right) \, ,\\
I^z_{q2} &=& - q^z v_z \gamma^3_z \left(\partial_t v_z
+ v_z \partial_z v_z \right) \, ,\\
I^z_{q3} &=& \frac{1}{5} \left[ \gamma^2_z \left( v_z \partial_t \pi + \partial_z \pi\right)
+  \gamma_z \pi \left(v_z \theta_z + \gamma_z \partial_t v_z \right)\right]
\end{eqnarray}
and
\begin{eqnarray}
I_{\pi 1} &=& \frac{1}{2}\pi
\left(\theta_z + D \ln \frac{\beta_2}{T} \right) \, ,\\
I_{\pi 2} &=& \frac{10}{9} (q^z\gamma^2_z)\left(\partial_t v_z
+ v_z \partial_z v_z \right) \, ,\\
I_{\pi 3} &=& \frac{2}{9} \left(v_z \partial_t q^z + \partial_z q^z
- \frac{q^z v_z}{\gamma_z} \theta_z \right) \, ,
\end{eqnarray}
where $q^z_{NS}$ and $\pi_{NS}$ are the Navier-Stokes values
\cite{Molnar:2009tx} and $D \equiv u^{\mu} \partial_{\mu}$.
$\tau_q$ and $\tau_\pi$ are the relaxation times, respectively.
For vanishing shear pressure and heat conductivity
the IS equations reduce to the equations of
ideal hydrodynamics.

\section{The relativistic Riemann problem}
\begin{figure}[th]
\includegraphics[width=15cm]{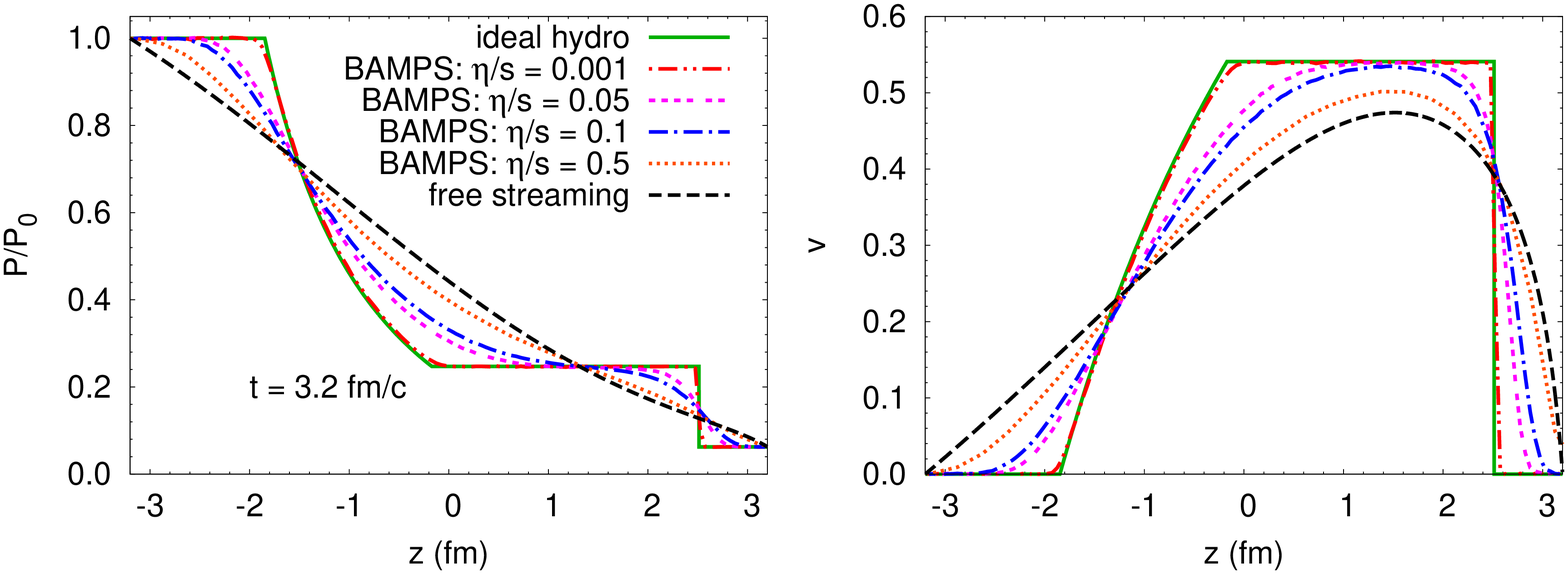}
\caption{Spatial profile of pressure (left panel, normalized by the initial
value) and collective velocity (right panel) at $t = 3.2 \, {\rm fm/c}$.
}
\label{fig:BAMPS_Shock}
\end{figure}

The relativistic Riemann problem \cite{Schneider:1993gd} is a well-known 
shock problem in ideal hydrodynamics.
Initially matter is separated by a membrane at $z = 0$ in two regions, $z<0$ and
$z>0$, with two different pressures $P_0$ and $P_4$. The velocities on
both sides are $v_0 = v_4 = 0$. The matter is assumed to be homogeneous
in transverse direction.\\
Removing the membrane in the case for a perfect fluid, i.e.,
$\eta/s = 0$, we observe (the green curve in Fig. \ref{fig:BAMPS_Shock})
a propagation of a shock wave to the right with a larger velocity than
the speed of sound and a rarefaction wave to the left exactly with the
speed of sound.

The BAMPS solutions for various $\eta/s$ are depicted in 
Fig. \ref{fig:BAMPS_Shock}. In particular, the BAMPS result for
$\eta/s = 0.001$ reproduces the ideal solution for a perfect fluid
to a very high precision. With the increasing $\eta/s$ value we see a clear 
transition from the formation of shock waves in ideal fluid to the smearing
out in free streaming of particles \cite{Bouras:2009nn}. The characteristic
shock profiles disappear for large $\eta/s$ values.

\begin{figure}[th]
\includegraphics[width=16cm]{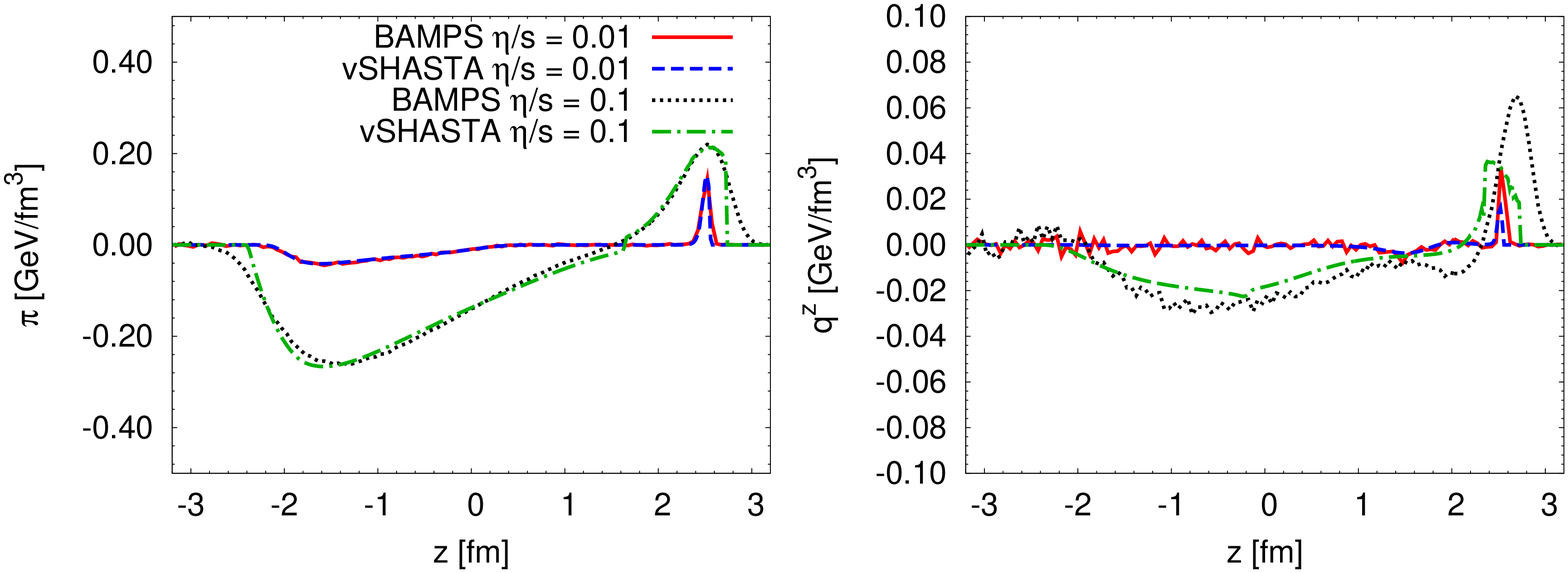}
\caption{Comparison of BAMPS and vSHASTA for the shear pressure $\pi$ and 
heat flow $q^z$.}
\label{fig:compare}
\end{figure}

In Fig. \ref{fig:compare} we show comparisions between BAMPS
and vSHASTA for the shear pressure $\pi = \pi^{zz} / \gamma^2$ and heat flow
$q^z = h \gamma^2 ( v N^0 - N^3)$ at $t = 3.2 \, {\rm fm/c}$.
$\pi^{\mu \nu}$ is the shear stress tensor, $N^\mu$ is the particle four-flow,
$\gamma^2 = (1-v^2)^{-1}$ and $h = (e + P)/n$ is the enthalpy per particle.
For $\eta/s = 0.01$ we see a very good agreement between vSHASTA and BAMPS,
whereas for $\eta/s = 0.1$ deviations in the region of the shock front
and rare faction fan appear. In the case of $\eta/s = 0.1$, the local system
at the shock front and rare faction fan is 
strongly out of equilibrium and, thus, the applicability of the IS theory 
is questionable. The microscopic transport approach does not suffer from
this drawback.

\section{Mach Cones in BAMPS}

\begin{figure}[th]
\begin{minipage}[l]{0.5\textwidth}
\includegraphics[width=8cm]{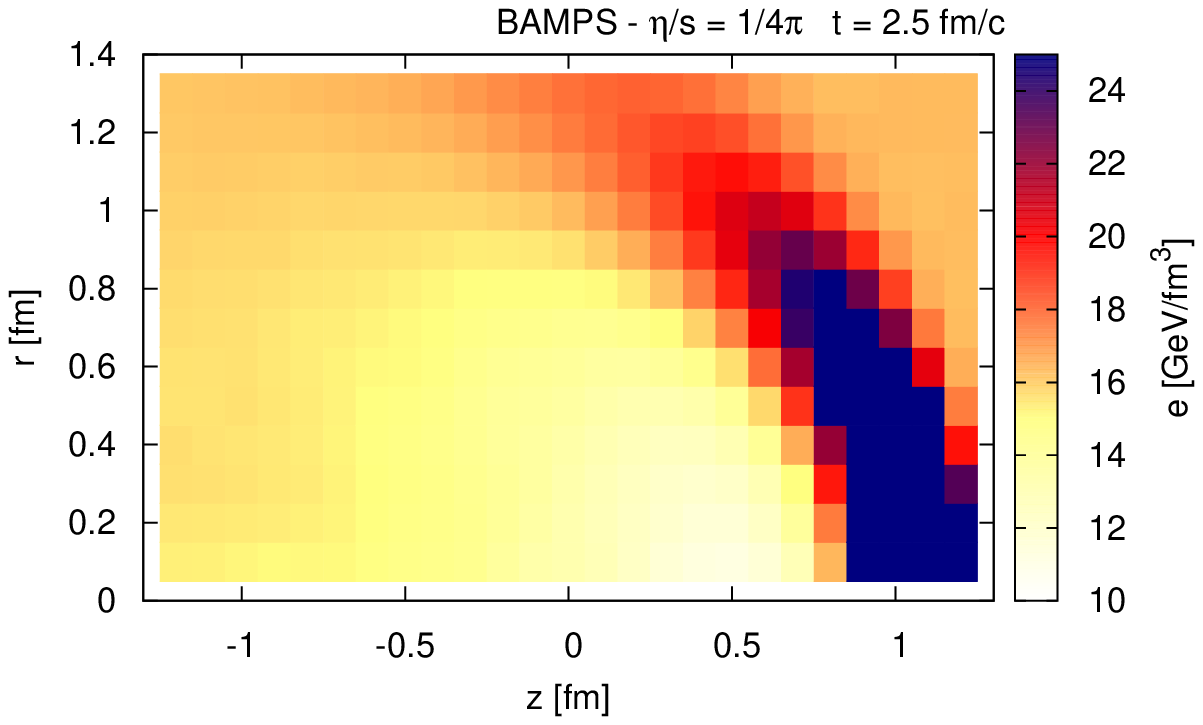}
\end{minipage}
\begin{minipage}[r]{0.5\textwidth}
\includegraphics[width=7cm]{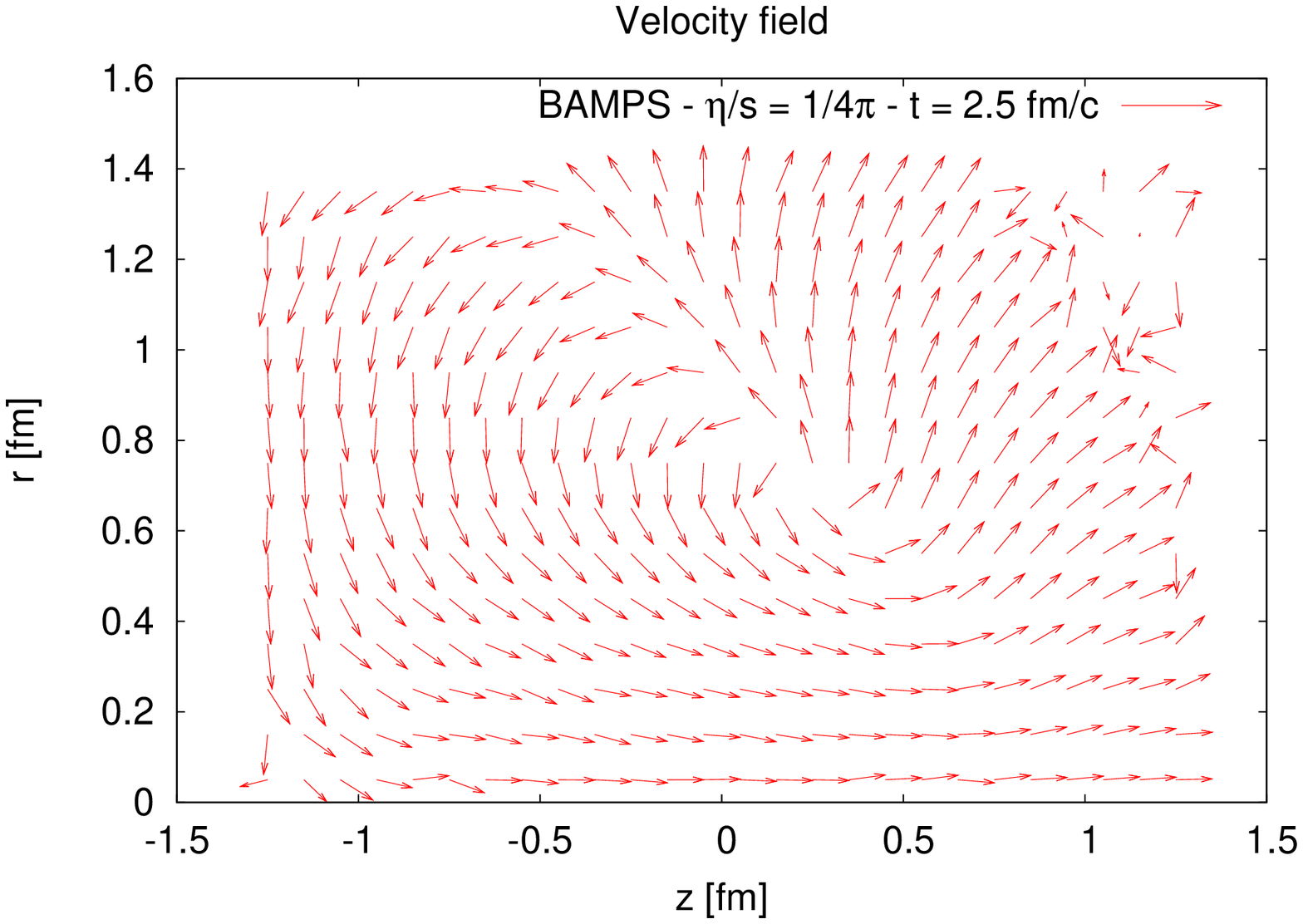}
\end{minipage}
\caption{Spatial profile of the energy density (left panel) and the velocity
field (right panel). The $\eta/s$ value of the thermal medium is $1/4\pi$.
The initial velocity of the high-energy gluon is
in z-direction. The length of the arrays in the right panel is unit, i.e.,
only the direction of the velocity is shown.}
\label{fig:machCone_0.08}
\end{figure}
Formation of 3-dimensional shock waves in form of Mach cones is investigated
by shooting a gluon with energy of $20 \, {\rm GeV}$ into a thermal gluon
medium with a temperature of $T = 400  \, {\rm MeV}$. The thermal
medium is embedded in a static box.

Fig. \ref{fig:machCone_0.08} shows spatial profiles of the energy 
density and velocity field at $t = 2.5 \, {\rm fm/c}$ for a medium 
with $\eta/s = 1/4\pi$. The energy, which the gluon probe lost due to
interactions with the medium, creates a shock wave that propagates
in form of a Mach cone. The energy density of the region
behind the Mach cone is smaller than the initial energy density of the medium.
This region is called a diffusion wake. Collective behavior of the medium
response is also clearly seen in the profile of the velocity field. 
Our results agree qualitatively with those found in \cite{Betz:2008ka,Molnar:2009kx}.
\begin{figure}[th]
\includegraphics[width=16cm]{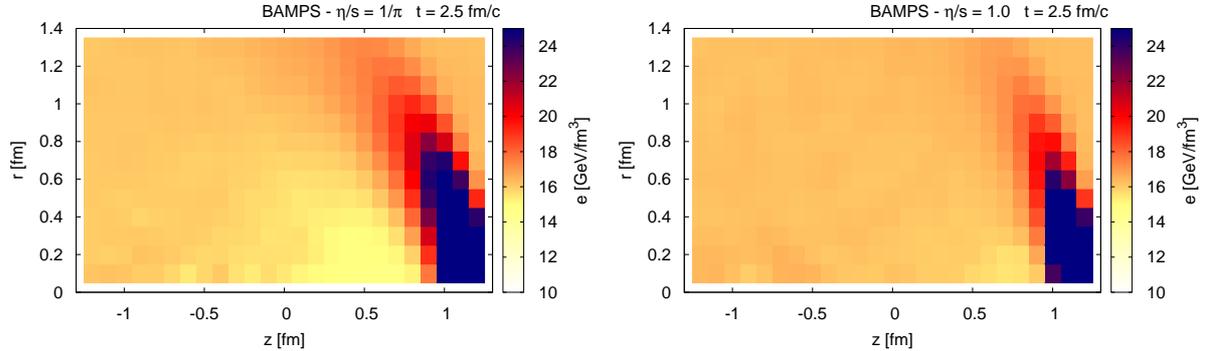}
\caption{Same as Fig. 3, but for $\eta/s = 1/\pi$
(left panel) and $1.0$ (right panel).}
\label{fig:machCone_visc}
\end{figure}
For higher values of the $\eta/s$ ratio the typical Mach cone structure
smears out as observed in Fig. \ref{fig:machCone_visc}.
The strength of the Mach cone signal and also the lower energy density
region behind the shock front become weaker because of weaker
particle interactions in medium with larger $\eta/s$.

\section{Summary}
\label{summary}
We have solved the relativistic Riemann problem using BAMPS and vSHASTA,
in order to investigate the formation of shock waves in viscous matter.
A transition from a characteristic shock profile in ideal fluid to
a complete smearing out in particle free streaming has been observed 
when increasing the $\eta/s$ value. For high $\eta/s$ values deviations
between BAMPS and vSHASTA occur, which indicates the break down of the
IS hydrodynamics. 
We have also investigated the formation of Mach cones within BAMPS.
For $\eta/s = 1/4\pi$ a Mach cone is clearly visible. For larger values
of $\eta/s = 0.32$ and $1.0$ the dissipative effect is so strong that
the typical Mach cone structure disappears.

\section*{Acknowledgements}

The authors are grateful to the Center for the Scientific 
Computing (CSC) at Frankfurt for the computing resources.
I. Bouras is grateful to HGS Hire.
E.\ Moln\'ar acknowledges the Return Fellowship support by 
the Alexander von Humboldt foundation. The work of H.\ Niemi was supported by
the Extreme Matter Institute (EMMI).

This work was supported by the Helmholtz International Center
for FAIR within the framework of the LOEWE program 
launched by the State of Hesse.

\section*{References}


\begin{thebibliography}{99}

\bibitem{Adams:2003kv}
  J.~Adams {\it et al.}  [STAR Collaboration],
  Phys.\ Rev.\ Lett.\  {\bf 91}, 172302 (2003);
  A.~Adare {\it et al.}  [PHENIX Collaboration],
  {\it ibid.} {\bf 101}, 232301 (2008).



\bibitem{Wang:2004kfa}
  F.~Wang  [STAR Collaboration],
  J.\ Phys.\ G {\bf 30}, S1299 (2004);
  J.~Adams {\it et al.}  [STAR Collaboration],
  Phys.\ Rev.\ Lett.\  {\bf 95}, 152301 (2005);
  S.~S.~Adler {\it et al.}  [PHENIX Collaboration],
  {\it ibid.} {\bf 97}, 052301 (2006);
  J.~G.~Ulery  [STAR Collaboration],
  Nucl.\ Phys.\  A {\bf 774}, 581 (2006);
  N.~N.~Ajitanand  [PHENIX Collaboration],
  {\it ibid.} {\bf 783}, 519 (2007);
  A.~Adare {\it et al.}  [PHENIX Collaboration],
  Phys.\ Rev.\  C {\bf 78}, 014901 (2008).



\bibitem{Stoecker:2004qu}
  H.~St\"ocker,
  Nucl.\ Phys.\  A {\bf 750}, 121 (2005);
  J.~Ruppert and B.~M\"uller,
  Phys.\ Lett.\  B {\bf 618}, 123 (2005);
  J.~Casalderrey-Solana, E.~V.~Shuryak and D.~Teaney,
  J.\ Phys.\ Conf.\ Ser.\  {\bf 27}, 22 (2005);
  V.~Koch, A.~Majumder and X.~N.~Wang,
  Phys.\ Rev.\ Lett.\  {\bf 96}, 172302 (2006).


\bibitem{Adler:2003kt}
  S.~S.~Adler {\it et al.}  [PHENIX Collaboration],
  Phys.\ Rev.\ Lett.\  {\bf 91}, 182301 (2003);
  J.~Adams {\it et al.}  [STAR Collaboration],
  {\it ibid.} {\bf 92}, 052302 (2004);
  B.~B.~Back {\it et al.}  [PHOBOS Collaboration],
  Phys.\ Rev.\  C {\bf 72}, 051901 (2005).


\bibitem{Luzum:2008cw}
  M.~Luzum and P.~Romatschke,
  Phys.\ Rev.\  C {\bf 78}, 034915 (2008);
  H.~Song and U.~W.~Heinz,
  arXiv:0812.4274.


\bibitem{Xu:2007jv}
  Z.~Xu, C.~Greiner and H.~St\"ocker,
  Phys.\ Rev.\ Lett.\  {\bf 101}, 082302 (2008);

\bibitem{Xu:2008av}
  Z.~Xu and C.~Greiner,
  Phys.\ Rev.\  C {\bf 79}, 014904 (2009).



\bibitem{Schneider:1993gd}
  V.~Schneider {\it et al.},
  J.\ Comput.\ Phys.\  {\bf 105} (1993) 92.

\bibitem{Xu:2004mz}
  Z.~Xu and C.~Greiner,
  Phys.\ Rev.\  C {\bf 71} (2005) 064901
  [arXiv:hep-ph/0406278].

\bibitem{Molnar:2009tx}
  E.~Molnar, H.~Niemi and D.~H.~Rischke,
  Eur.\ Phys.\ J.\  C {\bf 65} (2010) 615
  [arXiv:0907.2583 [nucl-th]].


\bibitem{Bouras:2009nn}
  I.~Bouras {\it et al.},
  Phys.\ Rev.\ Lett.\ 103 (2009) 032301

\bibitem{Xu:2007ns}
  Z.~Xu and C.~Greiner,
  Phys.\ Rev.\ Lett.\  {\bf 100}, 172301 (2008);

\bibitem{El:2008yy} 
  A.~El {\it et al.},
  Phys.\ Rev.\  C {\bf 79}, 044914 (2009)

\bibitem{Betz:2008ka}
  B.~Betz {\it et al.},	
  arXiv:0812.4401;

\bibitem{Molnar:2009kx}
  D.~Molnar,
  AIP Conf.\ Proc.\  {\bf 1182}, 791 (2009)
  [arXiv:0908.0299 [nucl-th]].

\end{thebibliography}

\end{document}